\documentclass{PoS}

\title{A low mass pixel detector upgrade for CMS}

\ShortTitle{A low mass pixel detector upgrade for CMS}

\author{\speaker{Hans-Christian  K\"astli}%
         \thanks{On behalf of the CMS pixel community.}\\
        Paul Scherrer Institut\\
        E-mail: \email{hans-christian.kaestli@psi.ch}}

\abstract{The CMS pixel detector has been designed for a peak luminosity of $1\times 10^{34}cm^{-2}s^{-1}$ and a total dose corresponding to 2 years of LHC operation at a radius of $4~{\rm cm}$ from the interaction region. Parts of the pixel detector will have to be replaced until 2015. The detector performance will be degraded for two reasons: radiation damage
of the innermost layers and the planned increase of the LHC peak luminosity by a factor of 2-3.
Based on the experience in planning, constructing
and commissioning of the present pixel detector,
we intend to upgrade the whole pixel detector in
2015. The main focus is on lowering the material
budget and adding more tracking points.
We will present the design of a new low mass
pixel system consisting of 4 barrel layers and 3
end cap disks on each side. The design comprises
of thin detector modules and a lightweight
mechanical support structure using CO$_2$ cooling.
In addition, large efforts have been made to
move material from the services out of the 
tracking region.}

\FullConference{9th International Conference on Large Scale Applications and Radiation Hardness of Semiconductor Detectors\\
                 30 September - 2 October 2009\\
                 Florence, Italy}

\begin{document}

\section{Context of the upgrade}
\begin{figure}[htb]
\begin{centering}
 \includegraphics[width=0.49\textwidth]{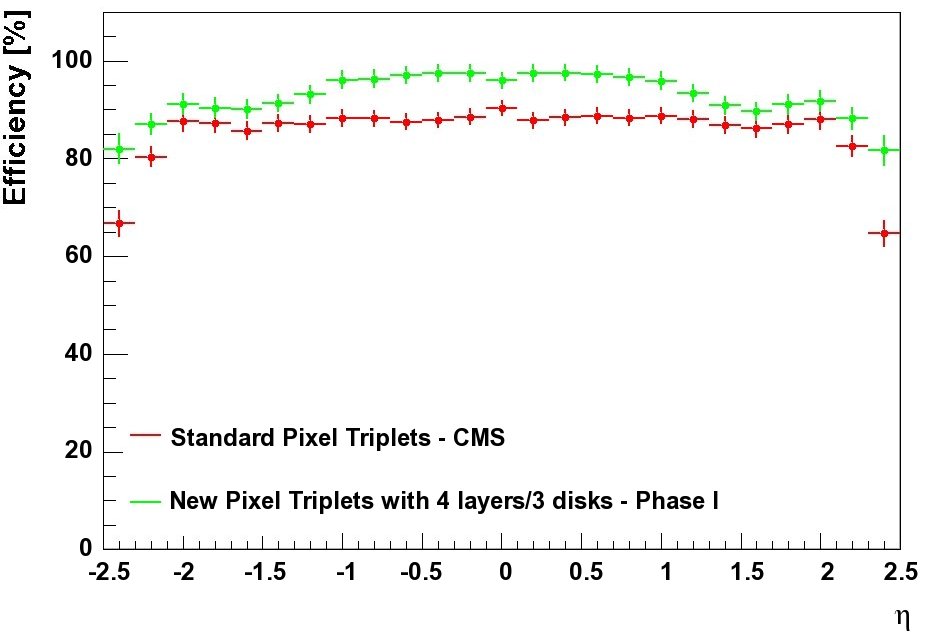}
 \includegraphics[width=0.49\textwidth]{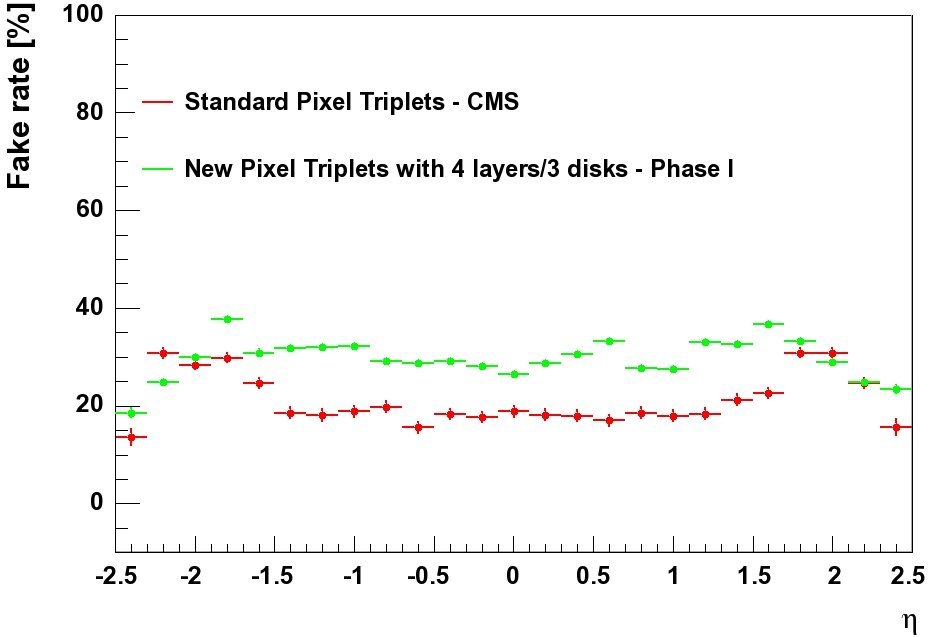}
    \caption{Preliminary comparison of the track seeding efficiency (left) and fake rate (right) in the pixel detector for $t\bar{t}$ events with the present algorithm (red) and with the addition of the 4$^{\rm th}$ barrel layer and 3$^{\rm rd}$ disks (green).}
    \label{fig:effi}
\end{centering}
\end{figure}

In the CMS experiment at the Large Hadron Collider (LHC) at CERN, the pixel detector is the tracking device closest to the interaction region \cite{bib:CMS}. It consists of three barrel layers and two end disks on each side. Due to its good 2 dimensional spatial resolution, it is best suited for track seeding in the high multiplicity environment and the measurement of secondary decay vertices from unstable particles.\\
In a first stage (phase I) the LHC upgrade program foresees an increase of the peak instantaneous luminosity by a factor 2-3 until the year 2015. This would significantly decrease the efficiency and performance of the present detector. Furthermore, the innermost layers will need replacement due to radiation damage of the sensor modules. Therefore, an upgrade of the pixel system for CMS is proposed here.\\
This upgrade project has to comply with several very restrictive boundary conditions. The most importants are 
\begin{itemize}
    \item {\bf Physics performance.} This must be the leading argument for the upgrade. To fully benefit from the increased luminosity, the upgraded pixel detector should be at least as performant in the higher radiation environment, as the present detector. For the pixel detector this translates into track seeding efficiency and secondary vertex resolution. This depends on 4 aspects:
    \begin{enumerate}
        \item[(a)] Track inter-/extrapolation length, i.e. distance between measurement layers.
        \item[(b)] Multiple scattering. 
        \item[(c)] Single point detection efficiency. 
        \item[(d)] Single point position resolution.
    \end{enumerate}
    Physics performance simulations are still under study. Preliminary results have been obtained with a simplified geometric model. Figure \ref{fig:effi} shows the track seeding efficiency and the fake rate in the pixel detector with the present algorithm and with the addition of the 4$^{\rm th}$ barrel layer and the 3$^{\rm rd}$ disks for simulated $t\bar{t}$ events. In the central region the efficiency is about 10\% higher with the new detector while the increase in fake track seeds is moderate. While full GEANT simulations are now under study, we address the underlying aspects (a) and (b) in section 2 and aspect (c) in section 3. The sensor design, the pixel frontend amplifier and therefore the single point position resolution will basically stay unchanged.

    \item {\bf Mechanical envelope.} The Si strip tracker and likely also the beam pipe with its suspension will not be modified in phase I. Therefore one cannot change the mechanical envelope and the insertion rails for the pixels. It has to fit exactly into the present tracker. There is some freedom in dividing the volume between pixel barrel and end disks which will be exploited to increase the hit coverage.
 
    \item {\bf Services.} The routing of the services from the experimental cavern to the inside of the solenoid (patch panel PP1) has been an extremely challenging task. Cooling pipes, power cables and optical fibres are tightly stacked in complex formed trays finding its way through outer CMS. Changing pixel services or even adding additional ones is not conceivable. There is some freedom to change services from PP1 to the detector, although one has to minimize the working time in an activated environment. 

    \item {\bf Resources needs.} For the LHC phase II upgrade, CMS needs a completely new inner tracker including the pixel detector. That project will go beyond what is feasible today and needs a lot of R\&D and funding to succeed. Manpower and funding are limited and to some extend have to be shared by the phase I and II upgrade projects. The phase I proposal tries to maximize the physics benefit with a minimal amount of resources by reusing existing technologies and components as much as possible.

    \item {\bf Commissioning.} In 2015, the CMS experiment will be in a competitive physics situation. The physics output must not be delayed due to the commissioning of upgraded detector parts. This implies that there are no significant changes in the operation of the pixel detector. We propose therefore a system based on the present front end electronics with identical detector control links and all essential configuration parameters unchanged. The proposed changes in the front end and the upstream data links are fully transparent to the detector control system, meaning that all operational procedures, such as powering sequence, threshold trimming or run start/stop, will be unchanged. 

\end{itemize}

\section{Mechanical structures}

\begin{figure}[htb]
\begin{centering}
 \includegraphics[width=0.9\textwidth]{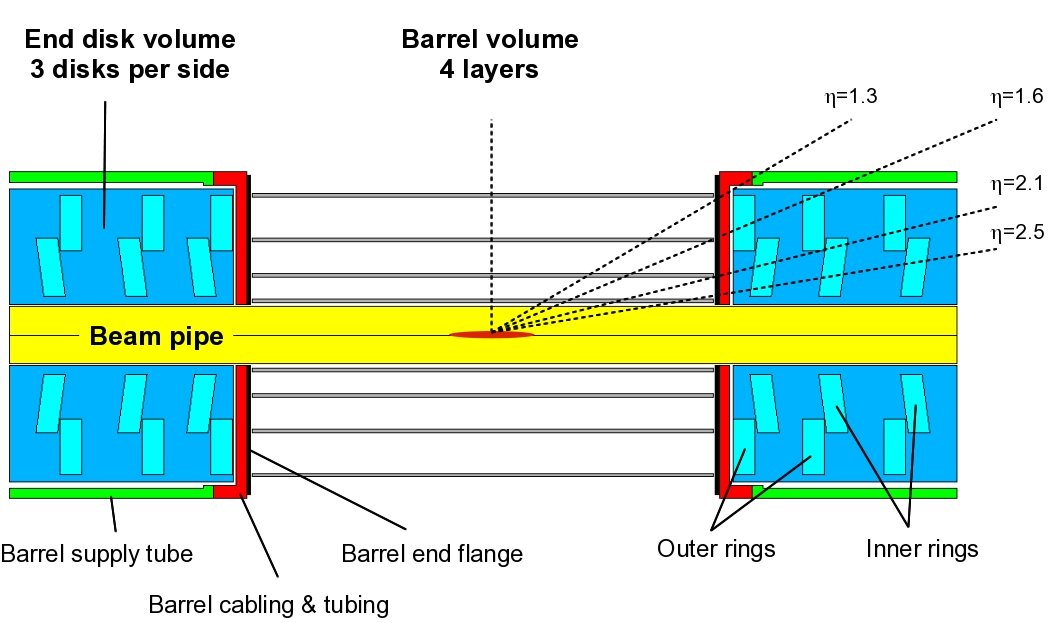}
    \caption{Schematic view of the upgraded pixel detector. It consists of four barrel layers and 3 end disk on each side. Each disk consists of an inner and an outer ring. Sensor modules of the inner ring are tilted outward by 20 degrees.}
    \label{fig:system}
\end{centering}
\end{figure}

Similar to the present pixel detector, the upgraded system consists of two half-barrels which are vertically separated with a left and a right set of half-disks on each side. The end disk sensor modules are mounted directly on the corresponding supply half-cylinders, while for the barrel the supply tubes are separate structures connected to the detector via low mass cables and cooling pipes. \\
A cross sectional view of the new system is shown in figure \ref{fig:system}. It consists of 4 barrel layers of equal length at radii 3.9, 6.8, 10.9 and $16.0~$cm and three end disks on each side, each one divided in an inner and outer ring. The radial coverage of the disks ranges from 4.5 to $16.1~$cm. The exact positions along the beam line are still to be defined. Shown are z=$\pm 35.9$, $\pm 39.6$ and $\pm 49.1~$cm.\\

{\noindent The focus of the design is threefold:}
\begin{enumerate}
    \item More robust tracking through 4 point coverage up to $\eta=2.5$ for tracks originating from $z=\pm5~{\rm cm}$ around the nominal center of the interaction region and a strongly reduced gap between the pixel detector and the rest of the Si tracker.
    \item Reduced material budget and hence multiple scattering and photon conversion rate. The largest gain comes from the change in the cooling system to a two phase CO$_2$ cooling and from the shift of end flange PCBs out of the tracking volume.
    \item Simpler design with only one type of module and fewer components.
\end{enumerate}

In this section we concentrate on the mechanical structures, while the electronics aspects are deferred to section 3.

\subsection{Pixel Barrel}

\begin{figure}[htb]
\begin{centering}
 \includegraphics[width=0.45\textwidth]{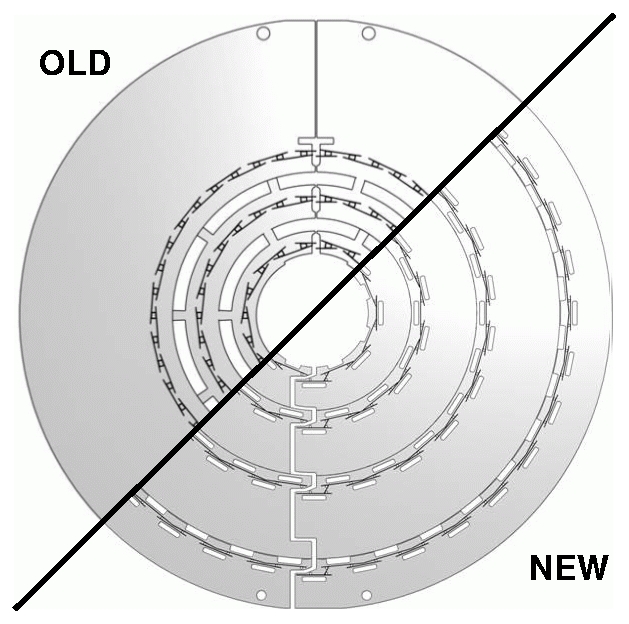}
    \caption{Cross sectional view of the old 3 layer barrel (top) and the new 4 layer system (bottom). The outermost layer drastically reduces the gap between the pixel detector and the Si strip tracker.}
    \label{fig:old-new}
\end{centering}
\end{figure}

A comparison of the old 3 layer barrel and the new 4 layer system is shown in figure \ref{fig:old-new}. 

\begin{figure}[htb]
\begin{centering}
 \includegraphics[width=0.6\textwidth]{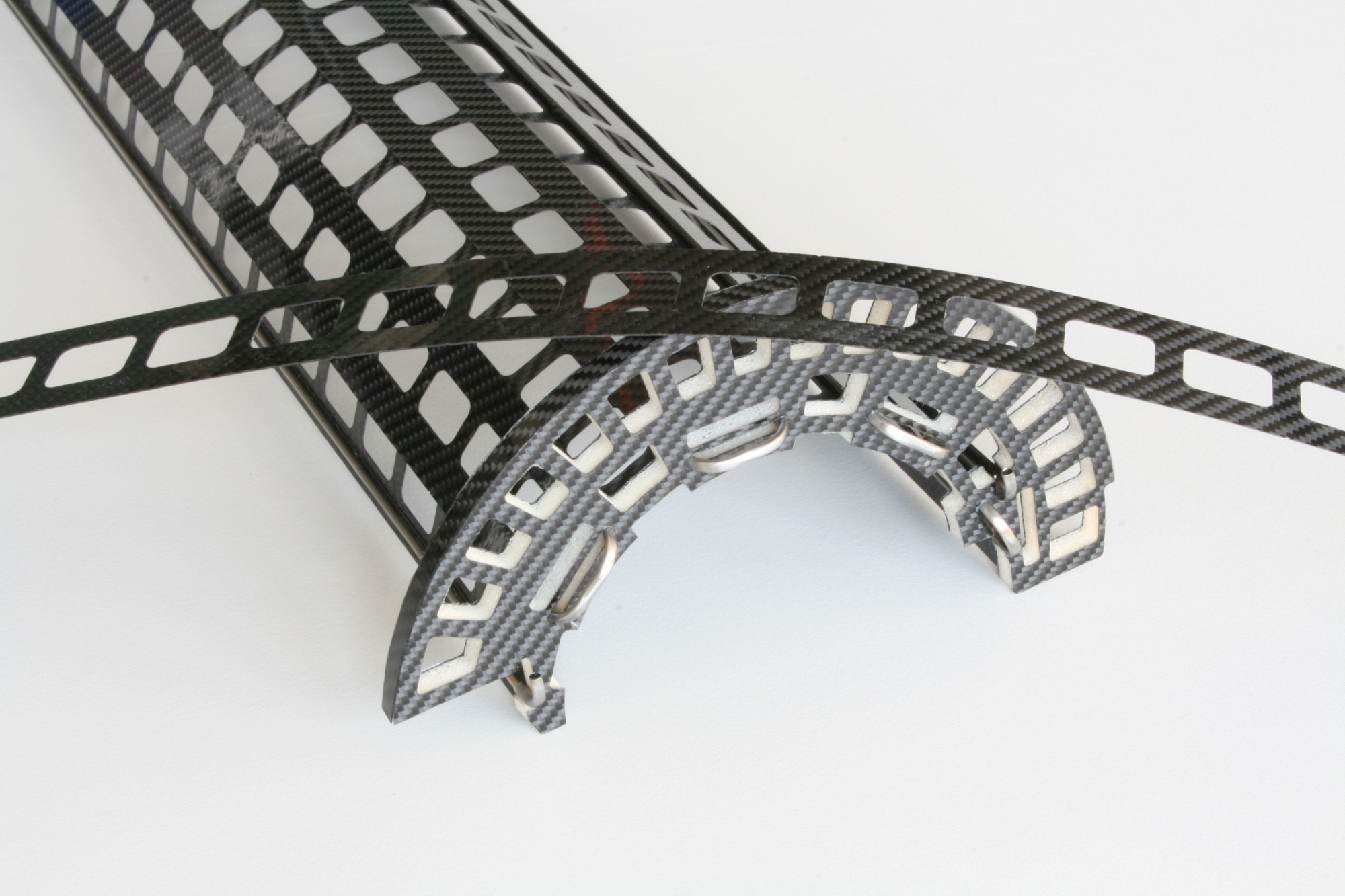}
    \caption{Prototype of the layer 1 mechanical structure. To illustrate the economic use of insensitive material, a carbon fiber ladder is laid upon the half-barrel. It bends like a sheet of paper. The mechanical stability comes only when glued to the cooling tubes.}
    \label{fig:barrel-mech}
\end{centering}
\end{figure}

A prototype of the 1$^{\rm st}$ barrel layer mechanics has been built at PSI. A picture of it is shown in figure \ref{fig:barrel-mech}. Emphasis has been placed on a minimal amount of material. It consists of $200~{\rm \mu m}$ carbon fiber ladders with cutouts to reduce the mass glued onto stainless steel cooling tubes with an outer diameter of $1.5~{\rm mm}$ and a wall thickness of $50~{\rm \mu m}$. Each ladder is glued to two tubes, alternating from above or below (see figure \ref{fig:old-new}). The end flange is made of the same carbon fiber material glued onto $4~{\rm mm}$ Airex foam profiles. The bends of the tubes are made from $1.8~{\rm mm}$ diameter $100~{\rm \mu m}$ thick stainless steel soldered to the straight section. It has been pressure tested up to $100~{\rm bar}$. The final weight of the structure is somewhat lower than estimated. With $51~$g it has 70\% less material per layer than the present design. Fully equipped with sensor modules, the new 4 layer barrel system will contain about 50\% less material than the present 3 layer system in the central region.

\begin{figure}[htb]
\begin{centering}
 \includegraphics[width=0.49\textwidth]{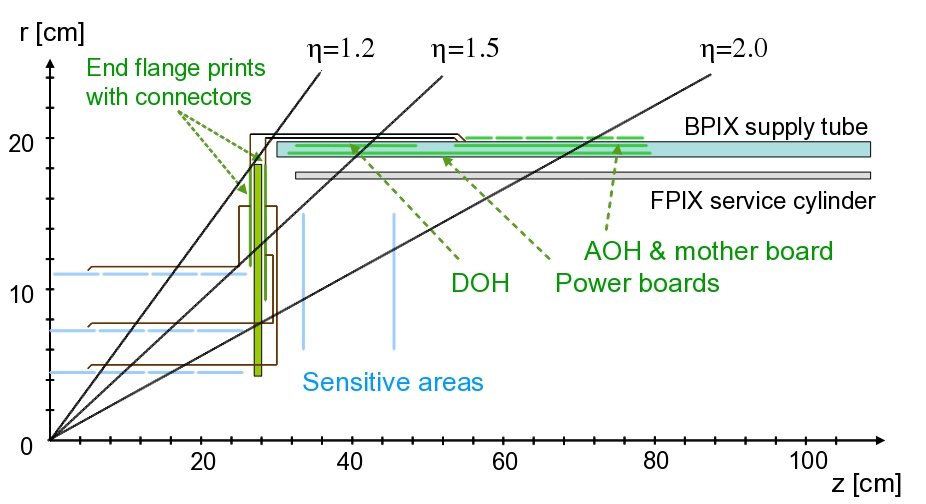}
 \includegraphics[width=0.49\textwidth]{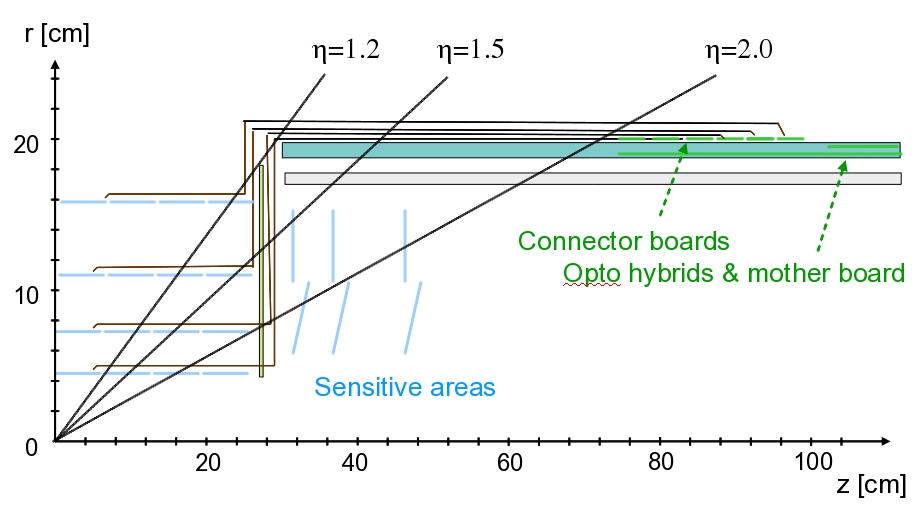}
    \caption{Illustration of the material savings in the range $1.2<\eta<2.0$. In the old design (left) there were PCBs with lots of connectors mounted on the end flange. Furthermore, tracks in this pseudorapidity range had to cross digital (DOH) and analog (AOH) opto hybrids with its mother- and daughter boards. In the new design (right) this has been moved to $\eta>2.0$.}
    \label{fig:supply}
\end{centering}
\end{figure}

Another important improvement is the shift of electronic components of the services to the forward direction. This is illustrated in figure \ref{fig:supply}. In the old design, PCBs were mounted on the detector end flange. They were equipped with lots of connectors where the module pig tail cables were attached. Tracks passing through this material suffer from multiple scattering, making the alignment, track and vertex reconstruction in the forward direction more difficult. Similarly, tracks with a pseudorapidity $\eta>1.5$ had to pass through several PCBs mounted on the supply tube. In the new design this has been substantially improved. The PCBs on the supply tubes are shifted to $\eta>2.0$ and those on the end flange have been removed entirely. This has been possible by developping new electrical links compatible with long low mass pig tail cables (see section 3). Modules are now connected directly to the analog opto hybrids (AOH) which are about $1~{\rm m}$ away.\\
A prototype design of the supply tube is finished and shown in \ref{fig:supply}. The supply tube is presently under construction.

\begin{figure}[htb]
\begin{centering}
 \includegraphics[width=0.8\textwidth]{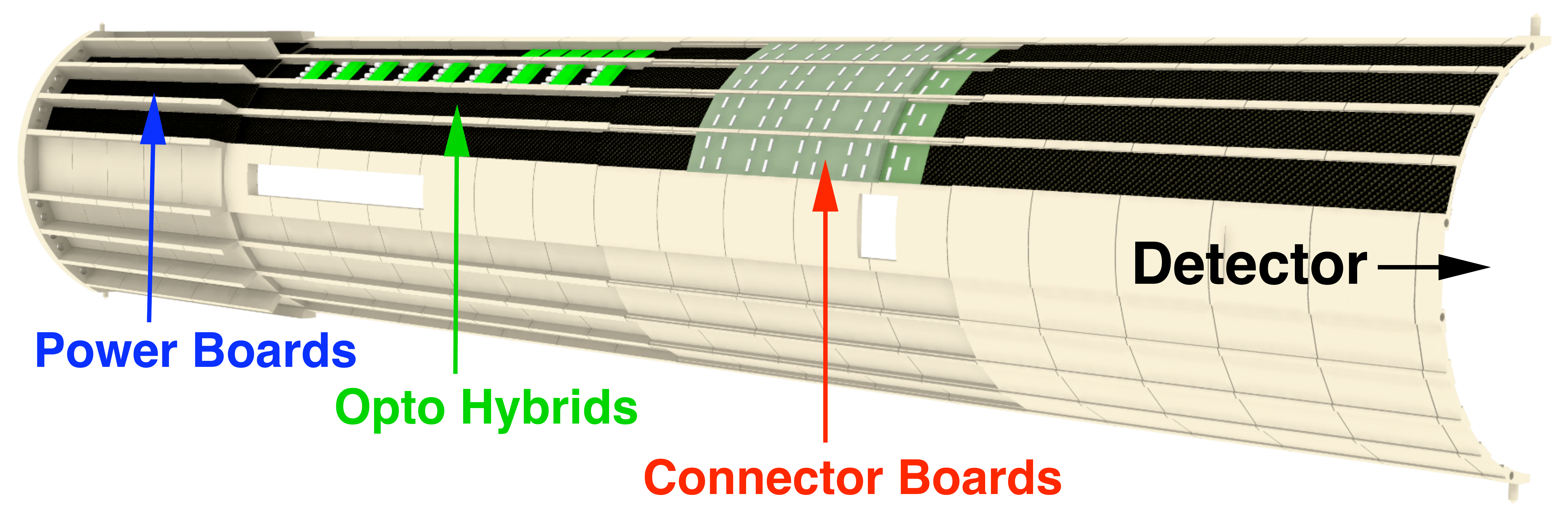}
    \caption{Drawing of the new supply half tube. Connector boards, opto hybrid and power boards are shifted away from the detector. Clearly visible is the empty region facing the detector. It consists only of a lightweight construction of carbon fiber ribs and Airex spacers. }
    \label{fig:supplytube}
\end{centering}
\end{figure}

\subsection{Pixel end disks}

\begin{figure}[htb]
\begin{center}
   \begin{minipage}[b]{0.25 \textwidth} 
      \includegraphics[width=0.7\textwidth]{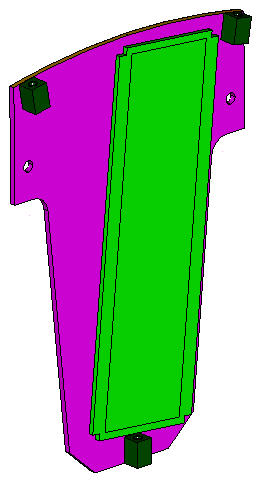}\\
      \includegraphics[width=0.7\textwidth]{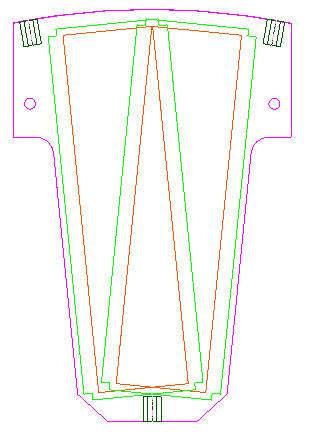}
   \end{minipage}
   \includegraphics[width=0.35\textwidth]{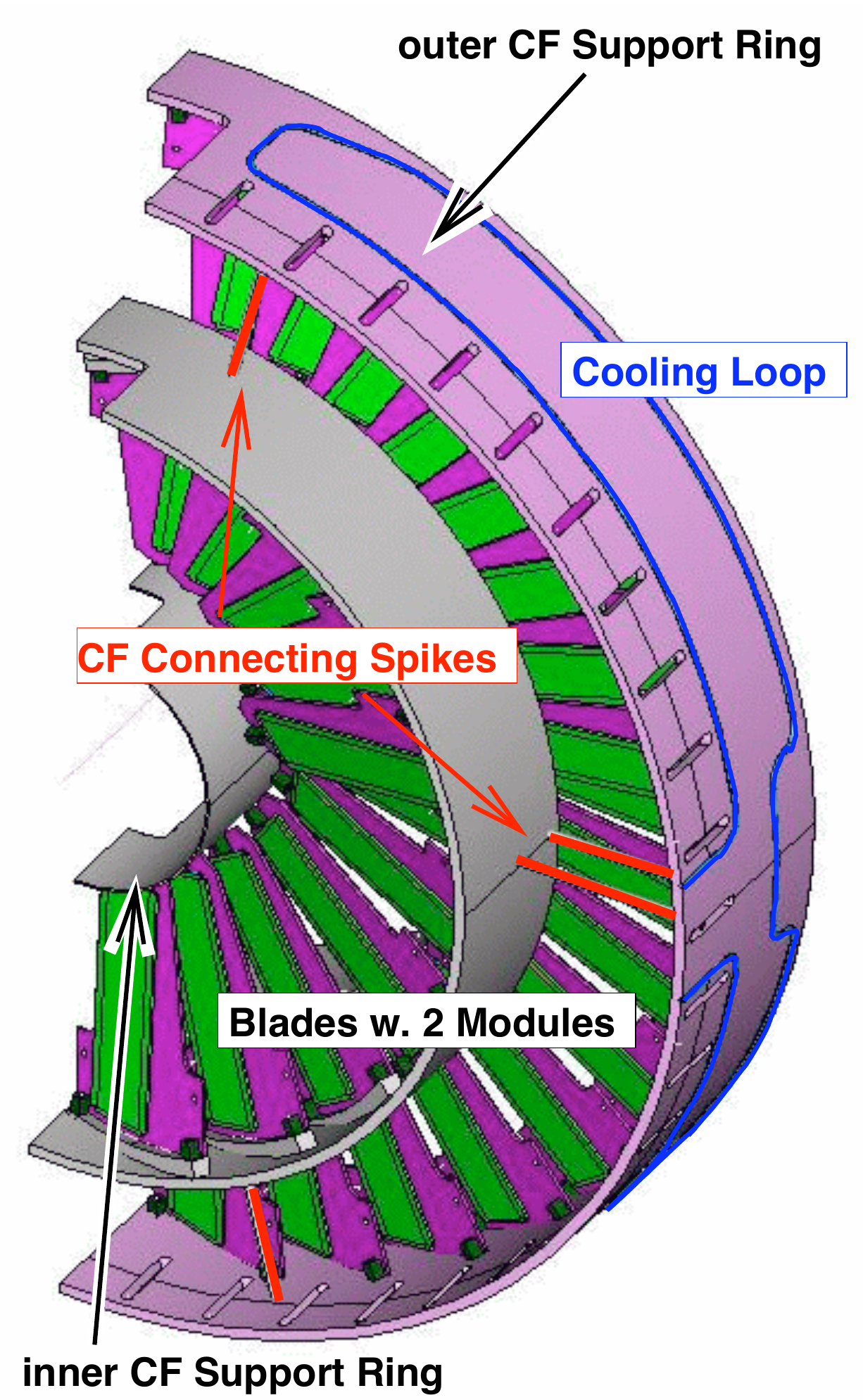}  
   \begin{minipage}[b]{0.25 \textwidth} 
     \begin{flushright}
      \includegraphics[width=0.7\textwidth]{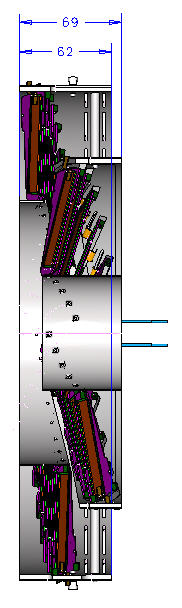}
     \end{flushright}
   \end{minipage}
   \caption{Drawings of the pixel end disk structure. The inner and outer rings consist of 11 and 17 blades respectively. All blades are identical (left pictures) and have two sensor modules (green) mounted on it, one on each side. They are mounted on two carbon fiber support rings (middle). The blades of the inner ring are tilted outward by 12 degrees in order to optimize hit coverage. All blades are rotated by 20 degrees around the radial axis to enhance charge sharing and hence position resolution. This is best visible in the side view (right picture).}
    \label{fig:half-disk}
\end{center}
\end{figure}

The mechanical structure of the end disks is more involved. Each of the 6 disks consists of two half-disks which in turn are made out of an inner and an outer half-ring. The structure holding the sensor modules is called a blade. The new design foresees only one type of blade shown in figure \ref{fig:half-disk} left. It is made of $0.88~{\rm mm}$ thermal pyrolytic graphite (TPG) with an excellent in plane thermal conductivity of $1500~{\rm W/mK}$ encapsulated in $0.06~{\rm mm}$ carbon fiber. A sensor module as described in section 3 is mounted on each side of the blade, giving some overlap of the sensitive area in phi. The outer (inner) ring consists of 17 (11) blades. All blades are rotated by 20 degrees around the radial axis to enhance charge sharing (same as in the present detector). In addition the blades of the inner ring are tilted outward by 12 degrees in order to optimize hit coverage with a minimal amount of sensor overlap. This is shown in the right part of figure \ref{fig:half-disk}.\\
Each ring has two carbon fiber support rings. Cooling tubes of $1.8~{\rm mm}$ outer radius and $100~{\rm \mu m}$ wall thickness are laid in grooves of the outer ring. The heat from the modules is transported radially through the TPG to this ring.

\section{Sensor Modules and Readout}

The new pixel detector has only one type of sensor modules. This will simplify considerably the testing and assembly\footnote{The present detector has 7 different type of plaquettes in the disks and full and 2 types of half modules in the barrel}. The module is based on the present barrel module. Changes are needed to reduce material budget and to cope with the higher data volume in phase I.

\subsection{Module design}

\begin{figure}[htb]
\begin{centering}
 \includegraphics[width=0.32\textwidth]{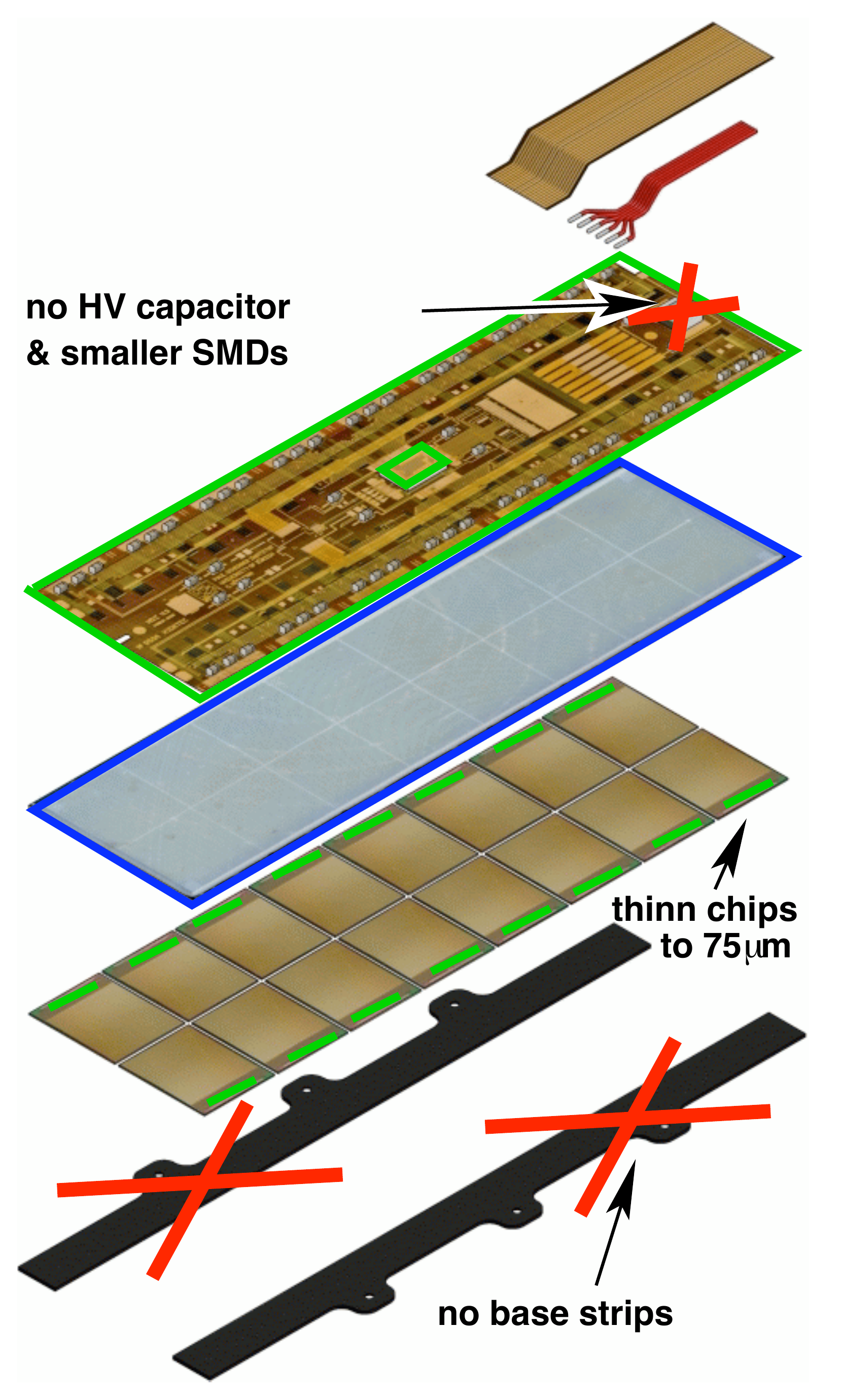}
    \caption{}
    \label{fig:module}
\end{centering}
\end{figure}

\begin{figure}[htb]
\begin{centering}
 \includegraphics[width=0.8\textwidth]{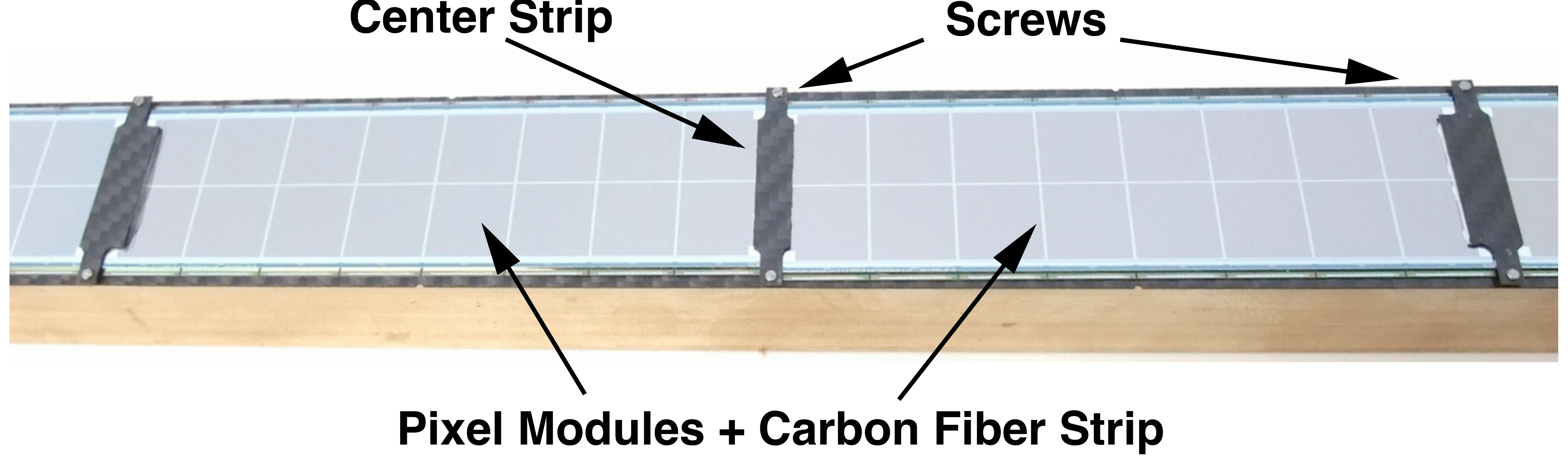}
    \caption{Carbon strips to mount the new base strip-less modules. A strip is glued on one module. It overlaps the second module fixing both positions when screwed onto the carbon structure.}
    \label{fig:Klammern}
\end{centering}
\end{figure}

Changes to the present barrel module design are illustrated in figure \ref{fig:module}. Like today, the module will consist of a sensor with an active area of $16.2\times64.8~{\rm mm^2}$ and 66560 $150~{\rm \mu m \times 100~\mu m}$ pixels. For the barrel the sensor will be unchanged (up to last minute material decisions). There is some ongoing R\&D activity in the forward pixel community to find optimized sensors. The sensor is bump bonded to $2\times8$ readout chips which, for barrel layer 1 and 2 are thinned down to $75~{\rm \mu m}$. A high density interconnect is glued on top of the sensor to distribute electrical signals. Based on experience with the present detector, some of the SMD components will be omitted, especially the relatively large (form factor 0612) high voltage filter capacitor. Power and Kapton signal cables are replaced by stranded low mass Cu cladded Al wires. The Si-nitride base strips are omitted to save material. A new base strip-less mounting procedure has been developed for the barrel, which is shown in figure \ref{fig:Klammern}. A small carbon fiber strip is glued to the end of a module. It overlaps the neighboring module and is screwed onto the mechanical structure using $500~{\rm \mu m}$ stainless steel screws. 

\subsection{Front end electronics}
The readout chip of the present detector (PSI46V2) is well understood and tested. Although modifications are needed, we want to reuse its core unchanged as much as possible. Changes are needed for two reasons:
\begin{enumerate}
    \item {\bf Single hit efficiency.} The chip has been designed for a peak luminosity of $10^{34}~{\rm s^{-1}cm^{-2}}$. All hit information has to be stored in internal buffers during the level 1 trigger latency (\cite{bib:hadi}). These buffers cannot cope with the increased luminosity of phase I. Therefore, data buffers must be increased. An additional chip internal buffer stage will be implemented which holds the level 1 trigger verified hits until a readout token arrives. This reduces dead time related to readout time and allows for a more efficient use of the link bandwidth. A software has been written to simulate the entire chip internal data flows and determine data loss rates. Pixel hit information are taken from full GEANT simulations. Table \ref{tab:sim} summarizes the total simulated data loss at the peak luminosity of $2\times10^{34}~{\rm s^{-1}cm^{-2}}$ and averaged over a fill assuming a luminosity lifetime of 10 hours and a turnaround time of the LHC beams of 5 hours. Most of the remaining inefficiencies can be compensated by the 4 hit coverage for track seeding as opposed to the present 3 hits.
    \item {\bf Faster readout links.} The total number of modules is 672 for the 6 disks and 1216 for the 4 barrel layers. In the barrel this is about $67\%$ more than today. There are not enough spare optical fibers and no additional fibers can be inserted. Therefore, new and faster links are needed. This is described in the next sub section. 
\end{enumerate}
The core of the chip with the pixel front end amplifier, threshold comparator with trimming and the column drain architecture stays unchanged. Also the downlink protocol for control and configuration of the detector is unchanged, both electrically and logically. This will considerably simplify the commissioning of the new detector. \\
The token bit manager chip \cite{bib:ed} is an integrated circuit designed to control a module. This chip will have to be redesigned as well. Like the readout chip, the core is basically unchanged. Only the data links will have to be changed to the new fast up-link protocol.

\begin{table}
\begin{center}
    \caption{Total electronics related inefficiency in \% at runstart with a peak luminosity of $2\times 10^{34}~{\rm s^{-1}cm^{-2}}$ and averaged over a LHC fill and a layer for the 4 barrel layers. For the average a luminosity lifetime of 10 hours and a turnaround time of 5 hours has been assumed.}
    \vspace{0.5cm}
    \begin{tabular}{|l|l|r|r|}
    \hline
       Barrel layer  & Radius [cm] & \multicolumn{2}{|c|}{Inefficiency [\%]}\\
       &  & peak & averaged \\
\hline
\hline
      1 & 3.9 & 4.7 & 2.1  \\
      2 & 6.8 & 0.7 & 0.38  \\
      3 & 10.9 & 0.1 & $<0.1$ \\
      4 & 16.0 & 0.1 & $<0.1$  \\
    \hline
    \end{tabular}
\label{tab:sim}
\end{center}
\end{table}

\subsection{Readout links}
\begin{figure}[htb]
\begin{centering}
 \includegraphics[width=0.35\textwidth]{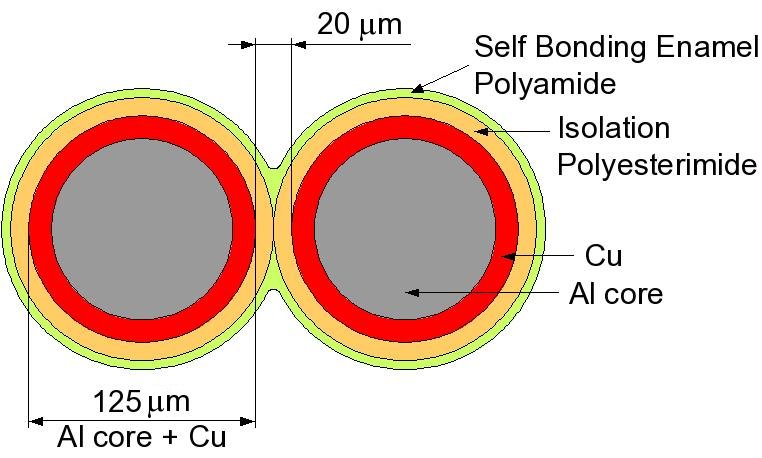}
 \includegraphics[width=0.5\textwidth]{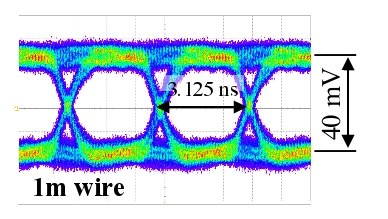}
    \caption{Left: Sketch of the low mass wire. The total thickness is $125~\mu$m. It has an Al core with $\approx 4-5\mu$m copper cladding. It is used as unshielded twisted pairs. Right: Eye diagram of the receiver signal for 1 m wire and 40 mV differential signals at 320 MBit/s. There is clear separation of the bits.}
    \label{fig:wire}
\end{centering}
\end{figure}

The new up-links will transmit data at 320 MBit/s from the modules via a $1~{\rm m}$ long pigtail cable to the opto converter on the service tubes. A very low power electrical signalling has been developed and special Cu cladded Al wires (\cite{bib:wire}) are used to reduce the mass. \\
The critical building blocks are an 8 bit on-chip analog-to-digital converter running at 80 MHz, digital transmitter/receiver running at 160/320 MHz and a PLL clock multiplier which generates the higher frequencies from the base 40 MHz which is distributed across the module. Prototypes of all these components have been designed and tested. The complete link has been tested at 320 MBit/s (see figure \ref{fig:wire}). The bit error rate is $<10^{-12}$ for 1 m cables at signal levels as low as 20 mV. This gives a total power per link of 4 mW or 12 pJ/Bit. More information can be found in \cite{bib:Beat,bib:Beat2}.

\section{Summary}
An upgrade of the CMS pixel detector is needed to cope with the higher luminosity foreseen in the LHC upgrade plans up to the year 2015. The focus of this upgrade is a drastic reduction of material budget and the adoption of the front end electronics and data links to the higher data rates. Existing components are reused as much as possible and the detector control is unchanged. This allows to increase the performance of the pixel detector with a minimal use of resources and commissioning time.

\end{document}